\documentclass[aps,12pt,final,notitlepage,oneside,onecolumn,nobibnotes,nofootinbib,superscriptaddress,noshowpacs,centertags]%
{revtex4}
\usepackage{graphicx}
\usepackage{txfonts}
\begin{document}
%\selectlanguage{russian} % Для  статьи на русском языке
%\selectlanguage{english} % Для статьи на английском языке
%
\title{Intergalactic electromagnetic cascade component \\ of observable blazar emission}
\author{\firstname{T.A.} \surname{Dzhatdoev}}
\email[]{timur1606@gmail.com}
\affiliation{Federal State Budget Educational Institution of Higher Education, M.V. Lomonosov Moscow State University, Skobeltsyn Institute of Nuclear Physics (SINP MSU), 1(2), Leninskie gory, GSP-1, 119991 Moscow, Russia}
\affiliation{Institute for Cosmic Ray Research, University of Tokyo, 5-1-5 Kashiwanoha, Kashiwa, Japan}
\author{\firstname{E.V.} \surname{Khalikov}}
\affiliation{Federal State Budget Educational Institution of Higher Education, M.V. Lomonosov Moscow State University, Skobeltsyn Institute of Nuclear Physics (SINP MSU), 1(2), Leninskie gory, GSP-1, 119991 Moscow, Russia}
\author{\firstname{A.P.} \surname{Kircheva}}
\affiliation{Federal State Budget Educational Institution of Higher Education, M.V. Lomonosov Moscow State University, Department of Physics, 1(2), Leninskie gory, GSP-1, 119991 Moscow, Russia}
\affiliation{Federal State Budget Educational Institution of Higher Education, M.V. Lomonosov Moscow State University, Skobeltsyn Institute of Nuclear Physics (SINP MSU), 1(2), Leninskie gory, GSP-1, 119991 Moscow, Russia}
\author{\firstname{E.I.} \surname{Podlesnyi}}
\affiliation{Federal State Budget Educational Institution of Higher Education, M.V. Lomonosov Moscow State University, Department of Physics, 1(2), Leninskie gory, GSP-1, 119991 Moscow, Russia}
\affiliation{Federal State Budget Educational Institution of Higher Education, M.V. Lomonosov Moscow State University, Skobeltsyn Institute of Nuclear Physics (SINP MSU), 1(2), Leninskie gory, GSP-1, 119991 Moscow, Russia}
\begin{abstract}
Secondary $\gamma$-rays from intergalactic cascades may contribute to observable spectra of blazars, also modifying observable angular and temporal distributions. In this paper we briefly review basic features of intergalactic electromagnetic cascade physics, suggest a new approximation for $\gamma$-ray mean free path, consider angular patterns of magnetically broadened cascade emission, and present an example of a fit to the observable blazar spectrum.
\end{abstract}
\maketitle
\section{Introduction \label{sec:intr}}

The nature of $\gamma$-ray emission from blazars ($\gamma$-ray loud active galactic nuclei) is still a mystery. Among many unsolved problems of blazar astrophysics one may name a role of intergalactic interactions of primary $\gamma$-rays in the formation of the observable spectrum. These $\gamma$-rays absorb on extragalactic background light (EBL) and cosmic microwave background (CMB) photons by means of the $\gamma \gamma \rightarrow e^{+}e^{-}$ pair production (PP) process \cite{nikishov62}--\cite{jelley66}. A secondary (cascade) observable $\gamma$-ray component may emerge due to inverse Compton (IC) scattering $e\gamma \rightarrow e^{'}\gamma^{'}$ \cite{jelley66}--\cite{protheroe86} (in what follows we do not distinguish electrons and positrons unless it is necessary). 

For several decades the so-called ``absorption-only model'' (AOM) that accounts only for primary photon absorption on the EBL and CMB and adiabatic losses \cite{stecker93}--\cite{abramowski13} has dominated extragalactic $\gamma$-ray propagation studies. However, some tentative deviations (``anomalies'') from the AOM were found \cite{neronov12}--\cite{furniss15}. It is possible to interpret all these anomalies as the result of intergalactic electromagnetic (EM) cascade development (for this line of reasoning see \cite{dzhatdoev17a}--\cite{dzhatdoev17c}). Indeed, in \cite{dzhatdoev17a} we have shown that the excess of observed $\gamma$-rays at the highest energy bins with respect to the expected intensity in the AOM framework \cite{horns12} could result from the existence of an ``ankle'' in the observable spectrum at the intersection of the primary and cascade components; the same holds for the effect of \cite{furniss15} (see \cite{dzhatdoev17b},\cite{dzhatdoev17d}). The anomalies found in \cite{neronov12} and \cite{chen15} are naturally explained by intergalactic electromagnetic (EM) cascade development. For the case of the most distant sources considered in \cite{rubtsov14} no firm conclusion about the presence of the anomaly similar to \cite{horns12} could be drawn as the EBL model uncertainties appear to be significant at high redshift (see Fig. 1 of \cite{fitoussi17}). 

\begin{figure}[t]
\centerline{\includegraphics[width=6.0in]{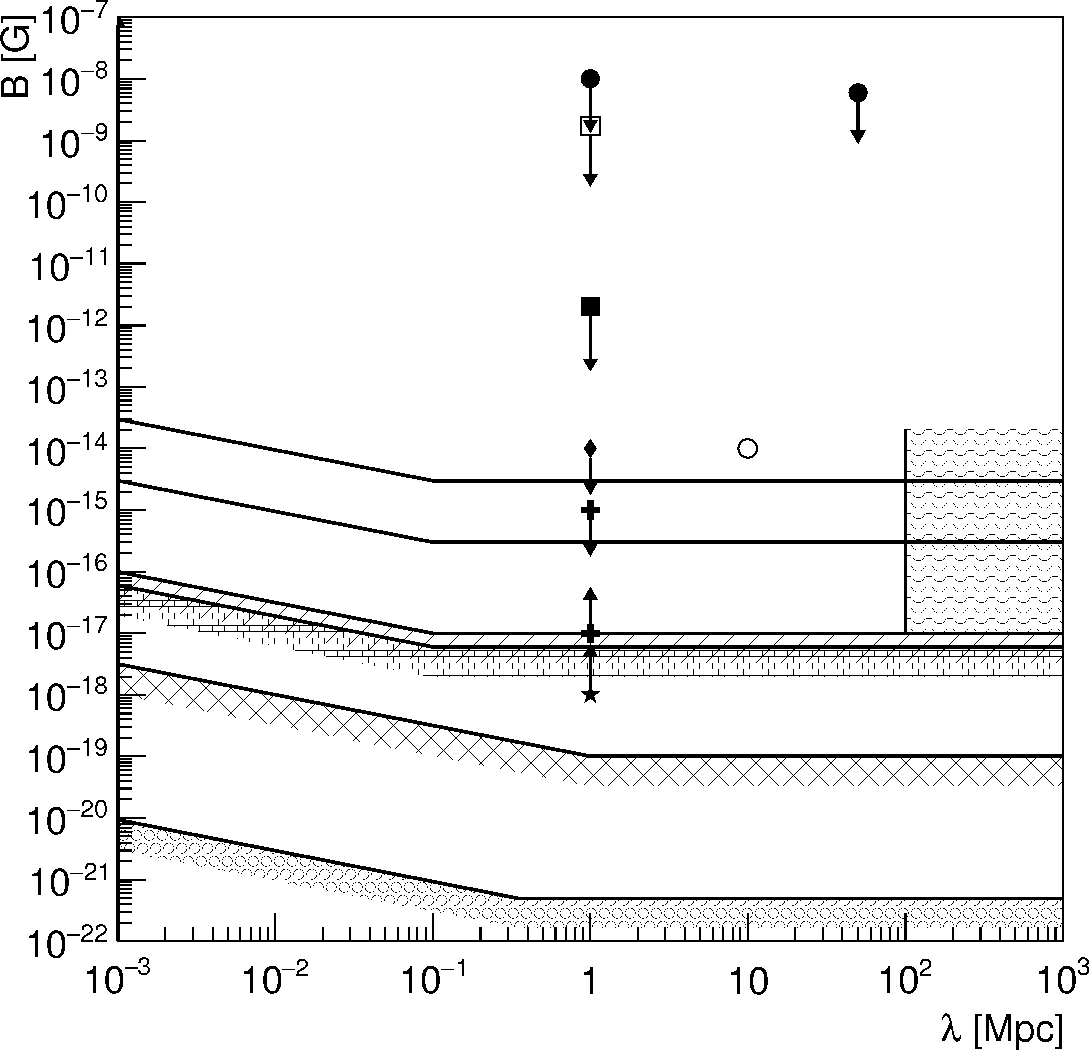}}
\caption{Some contemporary constraints on the EGMF strength depending on the coherence length (see text for details).}
\label{fig00}
\end{figure}

Observable properties of intergalactic EM cascades are heavily distorted by the so-called extragalactic magnetic field (EGMF). The current constraints on the strength of EGMF $B$ depending on the coherence length $\lambda$ are shown in Fig.~\ref{fig00}. In this picture arrows show directions to the regions where the EGMF strength does not contradict observations. Filled areas represent exclusion regions. Filled symbols mark the following constraints: circles --- \cite{blasi99}, square --- \cite{dolag05}, diamond --- \cite{prosekin12} (in this case it is assumed that the cascade component indeed contributes to the observable spectrum), plus signs - \cite{chen15}, star --- \cite{dermer11}, hollow square --- \cite{pshirkov16}. Hollow circle denotes the indication obtained in \cite{tashiro14}, bricks denote the constraint obtained by \cite{taylor11} and \cite{vovk12} for the EBL model of \cite{franceschini08}, vertical lines --- \cite{neronov10a}, slashed lines --- \cite{vovk12} (low limit from direct counts), horizontal lines --- \cite{abramowski14}, bubbles --- \cite{takahashi12}, waves --- \cite{tiede17}, criss-cross pattern --- \cite{finke15}.

A natural follow-up of our previous work would be the development of a unified model that could accomodate for all these anomalies simultaneously. This would require a good understanding of the effects imprinted by the EGMF on the observable spectrum, the angular and time distributions of observable $\gamma$-rays. To make a step towards this goal, in Sect.~\ref{sec:basic} we briefly describe the basic features of geometry and physics of intergalactic EM cascade. In Sect.~\ref{sec:mbc} we compute magnetically broadened cascade (MBC) angular patterns for certain conditions, compare their characteristics with angular resolution of various $\gamma$-ray instruments, and provide a fit to the observable spectral energy distribution (SED) of blazar 1ES 1218+304 with an account for the ``magnetic cutoff effect'' that diminishes the observable intensity at low energies. Finally, we draw out our conclusions in Sect.~\ref{sec:conclusions}.

In this work we consider only time-integrated observable spectra. We assume the EBL model of \cite{gilmore12}, unless stated otherwise.

\section{Geometry and physics of intergalactic EM cascade \label{sec:basic}}

In Subsect.~\ref{ssec:geometry} we recall the basic features of intergalactic EM cascade, in Subsect.~\ref{ssec:path} a new approximation for mean free path of $\gamma$-rays is provided, Subsect.~\ref{ssec:deflection} deals with deflection of cascade electrons while propagating through the EGMF.

\subsection{Basic considerations on the intergalactic cascade model \label{ssec:geometry}}

\begin{figure}[t]
\centerline{\includegraphics[width=6.0in]{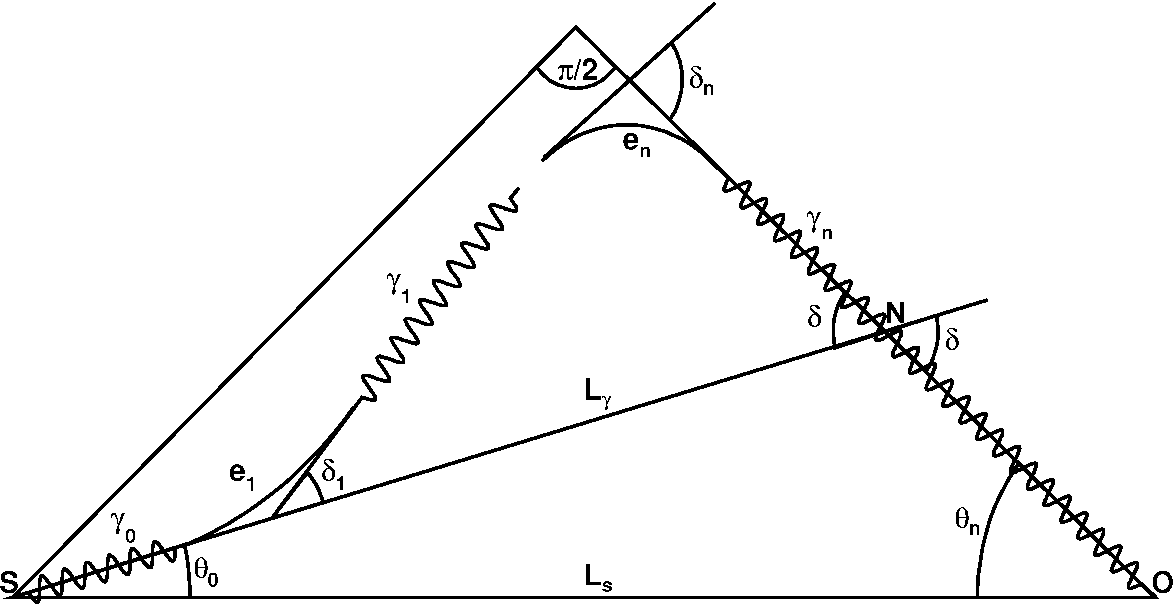}}
\caption{A sketch of intergalactic EM cascade geometry.}
\label{fig01}
\end{figure}

A basic sketch of an intergalactic EM cascade is presented in Fig.~\ref{fig01}. Let's assume that an individual $\gamma$-ray ($\gamma_{0}$) (primary energy $E_{0}$) was emitted by a source (S) at redshift $z_{s}$. $E_{0}$ is measured in the source restframe, neglecting any peculiar motions of the source. $L_{s}$ is the comoving distance from the source to the observer (O). The cascade may have one or more generations of electrons and $\gamma$-rays ($e_{1}$,$\gamma_{1}$), ($e_{2}$,$\gamma_{2}$) ,..., ($e_{n}$,$\gamma_{n}$) (only one of every two branches is shown for generation 1 and generation $n$ in Fig.~\ref{fig01}). $\gamma_{n}$ is an observable $\gamma$-ray; $\theta_{0}$ is the primary $\gamma$-ray emission angle with respect to the source-observer direction, $\theta_{n}$ --- the angle of the observable $\gamma$-ray with respect to the direction to the source. Cascade electrons of consecutive generations ($e_{1}$,$e_{2}$,...,$e_{n}$) acquire deflection angles ($\delta_{1}$,$\delta_{2}$,...,$\delta_{n}$) during the propagation through the EGMF; the angle between the rays SN and NO is denoted as $\delta$. In what follows we assume that ($\delta_{1} \ll \pi/2$, $\delta_{2} \ll \pi/2$ ,..., $\delta_{n} \ll \pi/2$), otherwise cascade electrons would be largely isotropised \cite{aharonian94}.

Electrons of each generation, on average, acquire energy $E_{e_{k}}= E_{\gamma_{k-1}}/2$; $k=1,2,...n$; $\gamma$-rays of each generation, on average, acquire energy
\begin{equation}
E_{\gamma_{k}}\approx (4/3)\epsilon \gamma_{e_{k}}^{2}, \label{eqn00}
\end{equation}
where $\epsilon$ is the mean energy of background photons, and $\gamma_{e_{k}}= E_{e_{k}}/m_{e}$ is electron Lorentz factor ($m_{e}$ is electron rest energy). Numerical estimates show that, as a rule, ($E_{e_{1}}\gg E_{e_{n}}$,$E_{e_{2}}\gg E_{e_{n}}$,...,$E_{e_{n-1}}\gg E_{e_{n}}$) and ($E_{\gamma_{1}}\gg E_{\gamma_{n}}$,$E_{\gamma_{2}}\gg E_{\gamma_{n}}$,...,$E_{\gamma_{n-1}}\gg E_{\gamma_{n}}$) \cite{aharonian02}, therefore:
\begin{eqnarray}
\delta_{k}<<\delta_{n}, k= 1,2,...,n-1 \label{eqn01}, \\
<L_{\gamma_{k}}> << <L_{\gamma_{n}}>, k= 1,2,...,n-1 \label{eqn02}, \\
<L_{e_{n-1}}> << <L_{\gamma_{n-1}}>, \label{eqn03}
\end{eqnarray}
where $<L_{\gamma_{k}}>$ is the $k^{th}$ generation $\gamma$-ray mean free path, and $<L_{e_{k}}>$ --- the $k^{th}$ generation electron mean free path. If the conditions (\ref{eqn01})-(\ref{eqn03}) are fulfilled, SN denoted as $L_{\gamma}$ in Fig. 1 may be approximated as $L_{\gamma}\approx L_{\gamma_{n-1}}$, and $\delta\approx \delta_{n-1}$ \cite{dolag09}. In what follows we assume that (\ref{eqn01})-(\ref{eqn03}) are fulfilled and call this approximation the dominating generation approximation. In the framework of this approximation $L_{\gamma}sin(\delta)=L_{s}sin(\theta_{n})$, or, again accounting for (\ref{eqn01})-(\ref{eqn03}):
\begin{equation}
sin(\theta_{n})= sin(\delta)\frac{L_{\gamma_{n-1}}(E_{\gamma_{n-1}},z_{s})}{L_{s}} \label{eqn04},
\end{equation}
similar to e.g. \cite{neronov07}--\cite{neronov09}.

\subsection{Approximation for mean free path \label{ssec:path}}

Here we propose a simple approximation for the mean free path of $\gamma$-ray $<L_{\gamma}(E,z)>$ (that is equal to the attenuation length) for the energy range 100 GeV$<E<$ 300 TeV and 0$<z<$1:

%\begin{equation}
%<L_{\gamma}(E,z)>= C\cdot L_{t} \frac{E_{t}}{(1+z)^{\alpha}E}[1+k\cdot sin(a\cdot lg[(1+z)^{\beta}E]-b)] \label{eqn05},
%\end{equation}
\begin{equation}
<L_{\gamma}(E,z)>= L_{att}(E,z)= C\cdot L_{t} \frac{E_{t}}{(1+z)^{\alpha}E}[1+k\cdot sin(a\cdot lg[(1+z)^{\beta}E]-b)] \label{eqn05},
\end{equation}
where $E_{t},L_{t}$ are fixed at (10 TeV, 80 Mpc), correspondingly, and the parameter values are $C$= 0.979, $\alpha$= 2.67, $\beta$= 0.857, $k$= 0.553, $a$= 3.04, $b$= 1.28. The values of parameters were estimated using the standard gradient minimization routine MINUIT \cite{james75} in the analysis framework ROOT \cite{brun97}. Here we utilized the approximate equality $L_{att}(E,z)\approx 1/R_{\gamma}(E,z)$, where $R_{\gamma}$ is the $\gamma$-ray interaction rate; cosmological corrections to $L_{att}(E,z)$ may be introduced following \cite{neronov09}.

\begin{figure}[t]
\centerline{\includegraphics[width=6.0in]{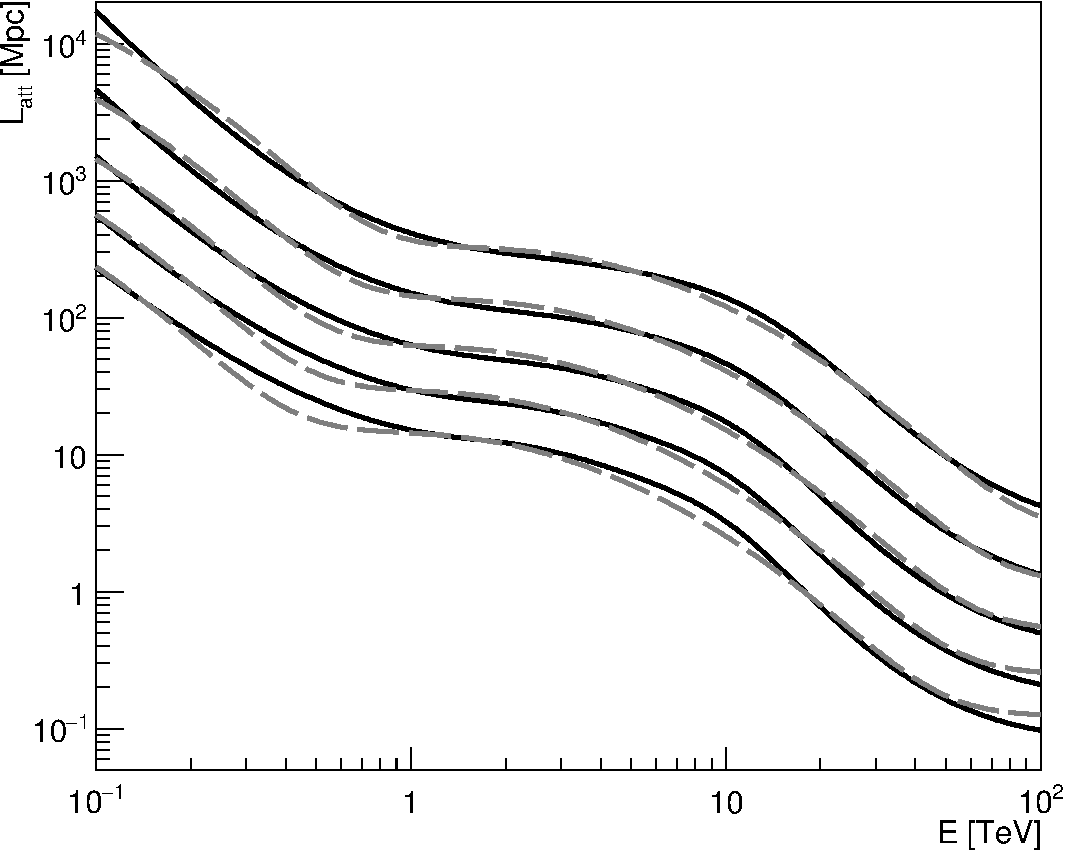}}
\caption{Mean free path (attenuation length) of $\gamma$-ray (solid black curves) and its approximation (dashed grey) for $z=(0,0.25,0.50,0.75,1.0)$ (from top to bottom; additional factors are introduced to better separate curves with different values of $z$).}
\label{fig02}
\end{figure}

\begin{figure}[t]
\centerline{\includegraphics[width=6.0in]{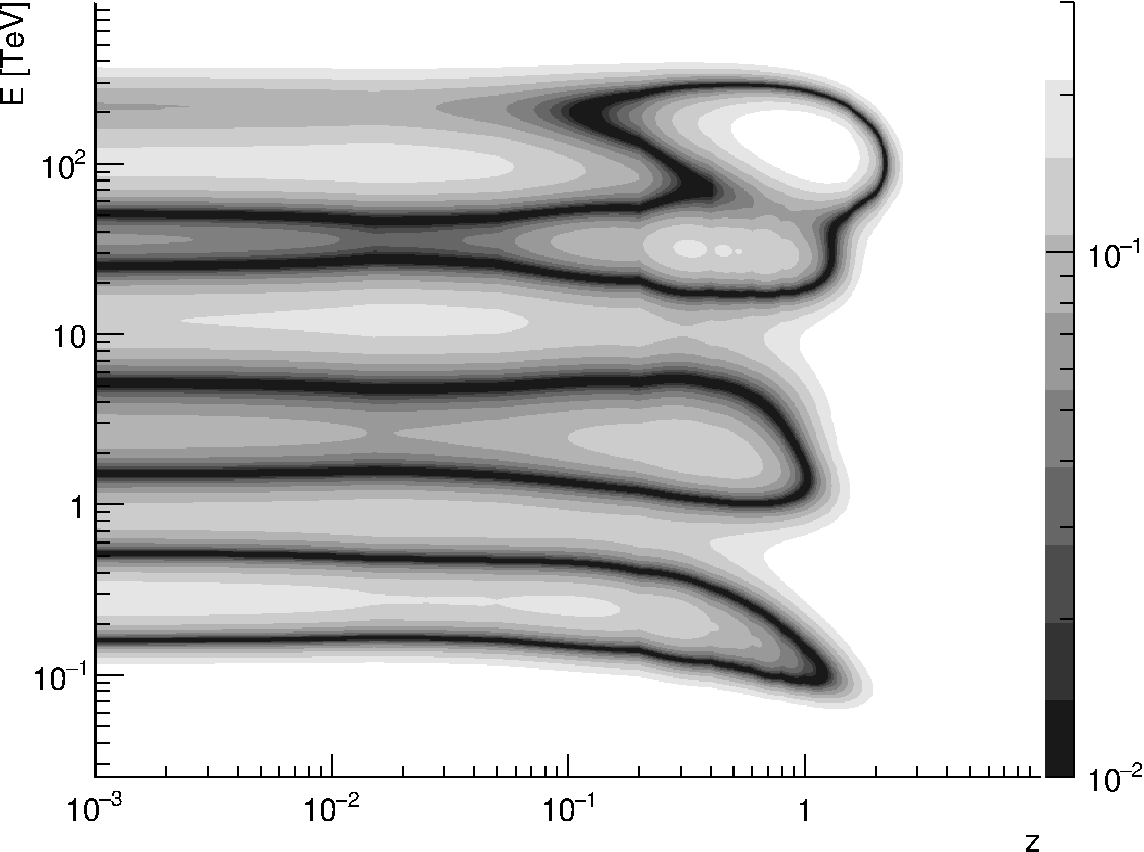}}
\caption{Relative accuracy of approximation (\ref{eqn05}) for $\gamma$-ray mean free path.}
\label{fig03}
\end{figure}

Numerical results for $L_{att}(E,z)$ together with its approximation are shown in Fig.~\ref{fig02}. Relative accuracy of this approximation is shown in Fig.~\ref{fig03} in grayscale. Values greater than 0.3 are shown in white, smaller than 0.01 --- in black.

\subsection{EGMF structure and electron deflection angle \label{ssec:deflection}}

In this paper we specialize on the case where the coherence length of the EGMF is $\lambda_{B}\sim$1 Mpc. For the range of parameters considered here, electron energy loss length $L_{E-e}$ is typically smaller than $\lambda_{B}$, but, on the other hand, $\lambda_{B}<<L_{\gamma_{n-1}}$, i.e., electrons lose most of their energy while propagating in coherent magnetic field, but the cascade develops over many EGMF cells. In this paper we consider the simplest case when the line-of-sight coincides with the axis of the primary $\gamma$-ray beam. Under these conditions, the MBC pattern appears to be practically axially symmetric \cite{neronov10b},\cite{broderick16}. Moreover, it is assumed that the evolution of the EGMF is simply dilution, i.e., $B(z)\sim(1+z)^{2}$ \cite{grasso01}. 

We propose an approximation for the deflection angle $\delta$ for 0$<z<$1 and electron energy (measured in the comoving frame at the pair-production redshift) 1 TeV$<E_{e}<$100 TeV. We assume the fixed distance between secondary (cascade) $\gamma$-ray production acts, $(1/(1+z)^{3})\cdot(1/R_{e})\approx$1.1 kpc$/(1+z)^{3}$, and that the energy of these cascade photons $E_{\gamma-obs}$ [GeV] (measured in the comoving frame at the production redshift) is described by (\ref{eqn00}), accounting for the corresponding electron energy loss. Therefore, $\delta$ depends on $E_{e}$, $z$, but also on $E_{\gamma-obs}$, and:
\begin{equation}
\delta(E_{e},z,E_{\gamma-obs})= C\frac{B(z)}{B_{t}}\cdot E_{\gamma-obs}^{-\alpha} \cdot e^{\frac{E_{t}}{E_{\gamma-obs}-E_{t}}} \label{eqn06},
\end{equation}
where $B$ is the EGMF strength, $E_{t}=K_{E}\cdot E_{c} \cdot E_{e}^{\beta}$ with $E_{c}$ from (\ref{eqn00}) at the electron's production point; $B_{t}=10^{-16}$ G = 0.1 fG, and the parameter values are $C= 0.150/(1+z)^{3}$, $\alpha$= 1.00, $\beta= 4.26 \cdot 10^{-2}$, $K_{E}$= 1.25. The values of these parameters were also estimated using MINUIT.

\section{Magnetically broadened cascade pattern and the magnetic cutoff effect \label{sec:mbc}}

\begin{figure}
\begin{minipage}{0.49\linewidth}
\centerline{\includegraphics[width=0.95\linewidth]{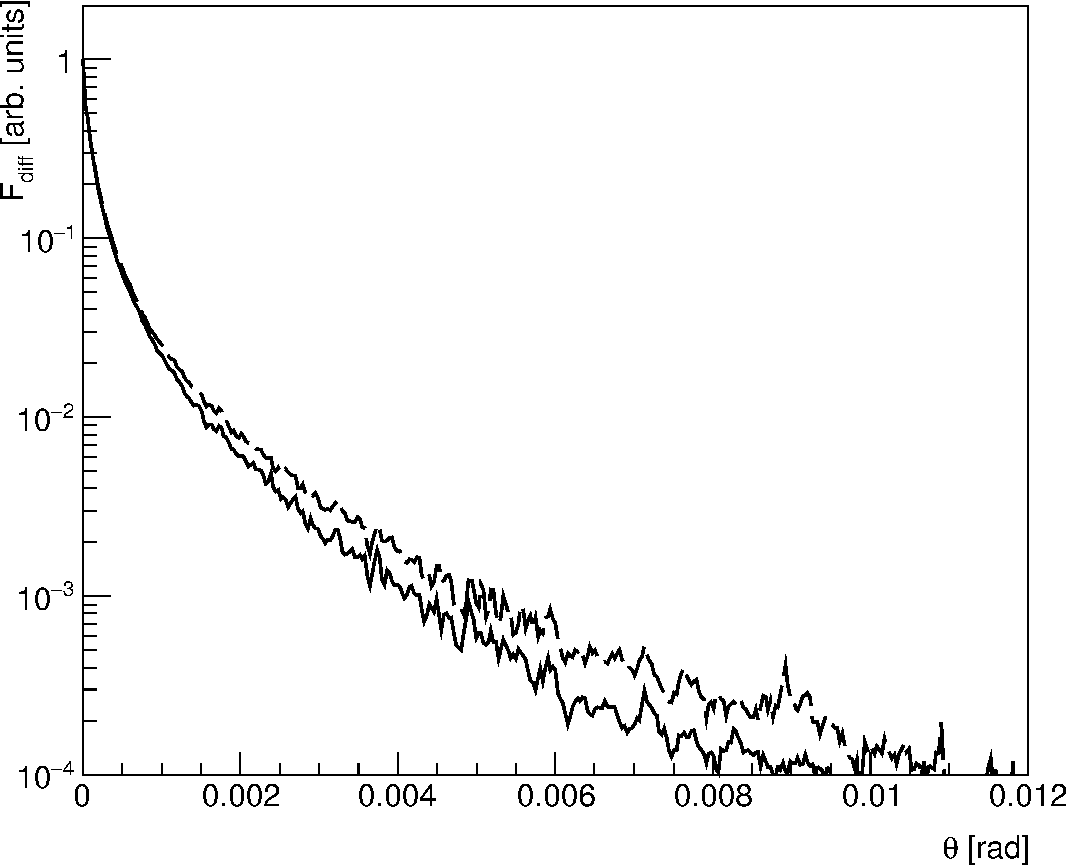}}
\end{minipage}
\hfill
\begin{minipage}{0.49\linewidth}
\centerline{\includegraphics[width=0.95\linewidth]{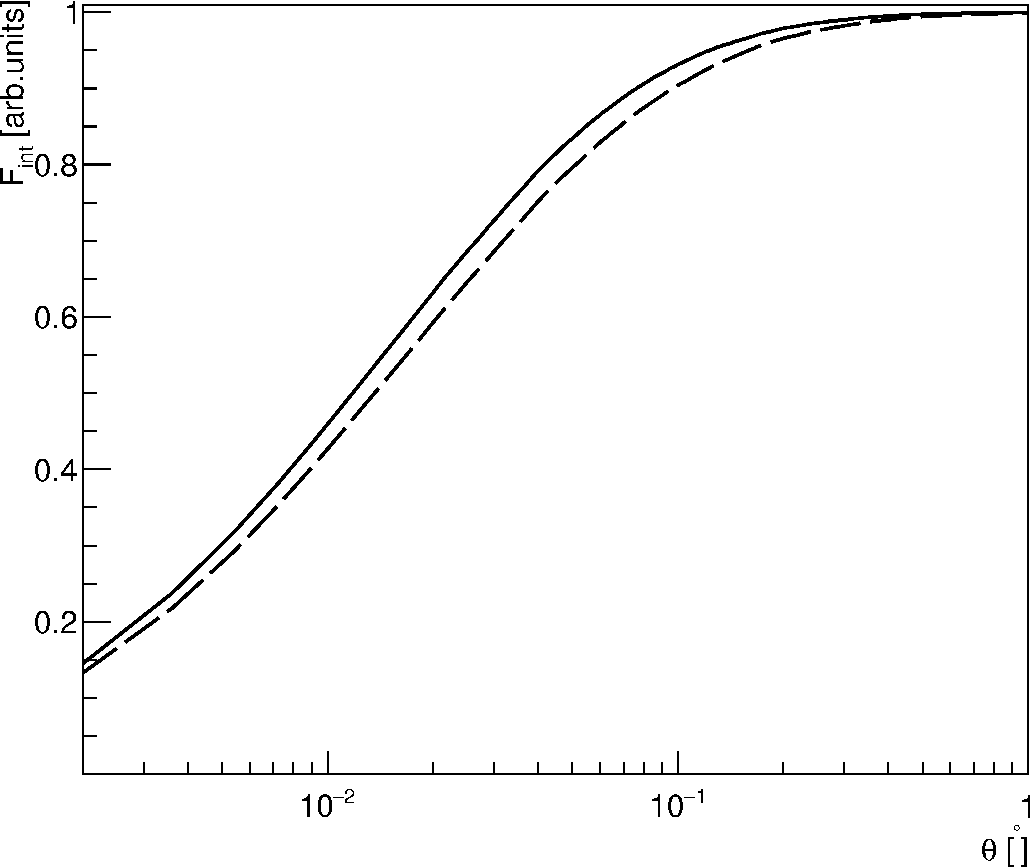}}
\end{minipage}
\newline
\begin{minipage}{0.42\linewidth}
\centerline{\includegraphics[width=0.95\linewidth]{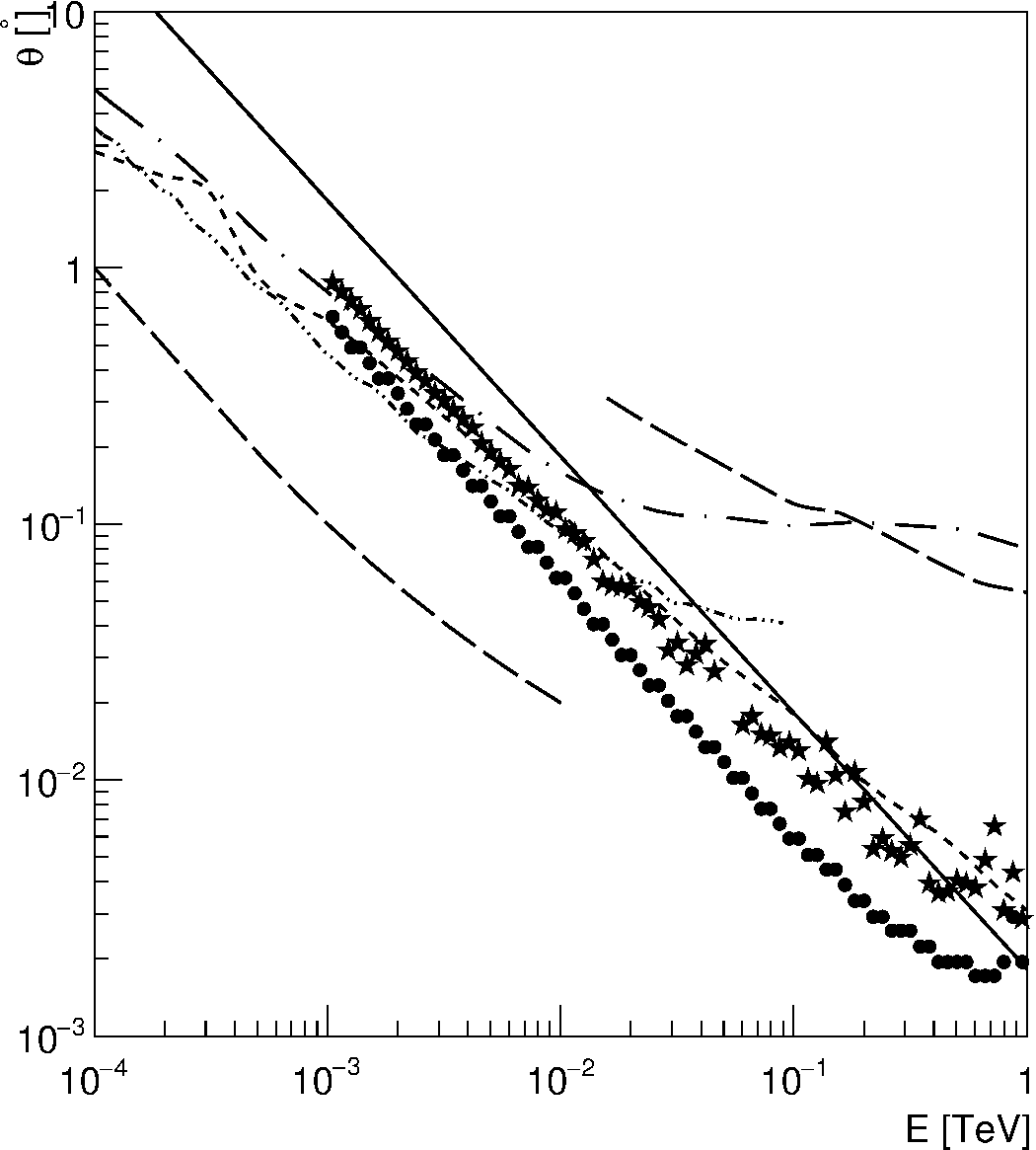}}
\end{minipage}
\hfill
\begin{minipage}{0.56\linewidth}
\centerline{\includegraphics[width=0.95\linewidth]{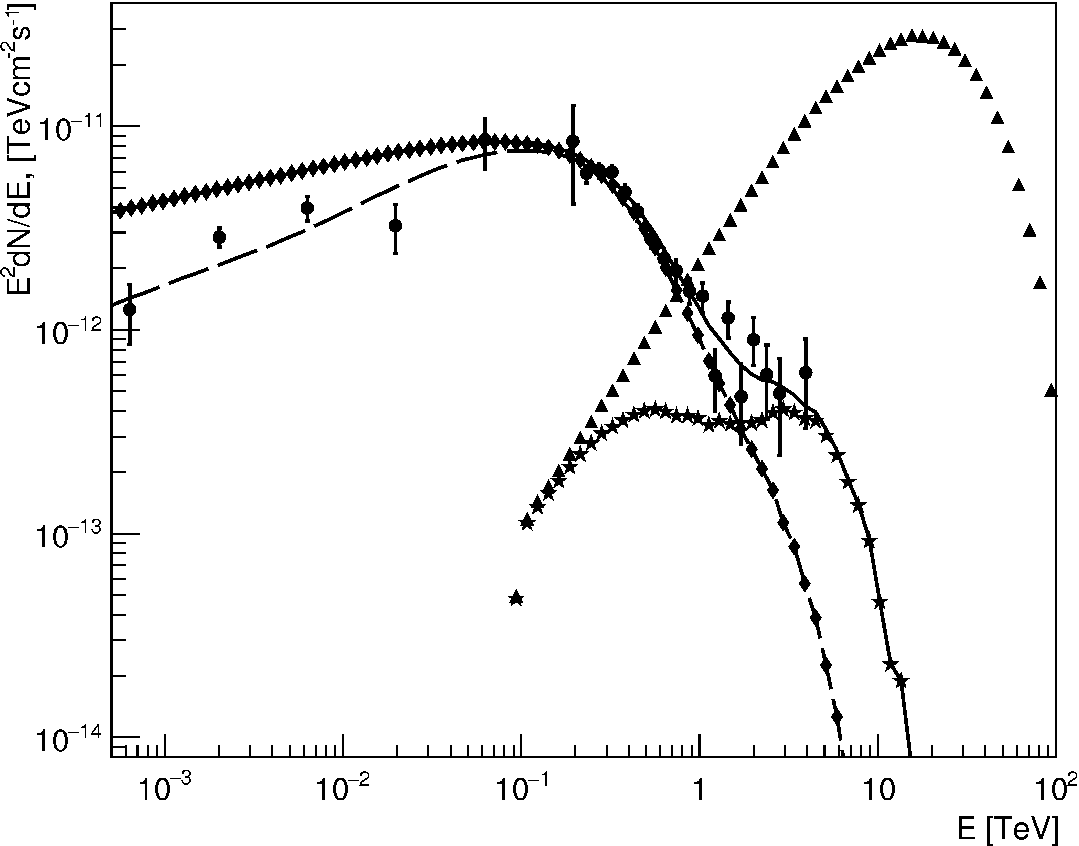}}
\end{minipage}
\caption[]{Top-left: the MBC pattern for the $\theta_{p}$= $1.0\cdot10^{-2}$ rad (solid curve) and $\theta_{p}$= 0.1 rad (dashed curve). Top-right: integral version of the same function for the same values of $\theta_{p}$. Lower-left: the MBC parameters compared to the angular resolution of various $\gamma$-ray instruments (see text for details).
Lower-right: a fit to the SED of blazar 1ES 1218+304 (see text for details).}
\label{fig04}
\end{figure}

We use the code of \cite{fitoussi17} to calculate the MBC pattern for some specific conditions (the primary $\gamma$-ray energy 10 TeV, observable $\gamma$-ray energy 10 GeV, redshift of the source $z_{s}$= 0.186, $B(z=0)$= 0.1 fG, the EGMF coherence length= 1 Mpc). In Fig.~\ref{fig04}, top-left, we present the MBC angular pattern calculated by us for various ranges of integration over the polar angle $\theta_{p}$ of observable $\gamma$-rays with respect to the direction to the source. We have checked that the dependence of the result over the range of $\theta_{p}$ = 0.03--0.3 rad is negligible. The integral version of the MBC pattern is shown in Fig.~\ref{fig04}, top-right.  

Fig.~\ref{fig04} lower-left presents the comparison of the MBC integral parameters vs. energy with similar quantities for various $\gamma$-ray instruments. The 68 \% containment angle $\theta_{68}$ for the CTA array's point spread function (PSF) \cite{acharya13} is shown by long-dashed curve at $E>$ 20 GeV, for the Fermi LAT telescope according to \cite{atwood09} --- by dash-two-dotted curve. The same quantity for the case of the new version of the Fermi LAT PSF (current version Pass8R2\_V6) is denoted by long-dash-dotted curve, for the GAMMA-400 instrument \cite{galper13} --- by short-dashed curve, and for the GRAINE emulsion $\gamma$-ray telescope \cite{takahashi15} --- by long-dashed curve up to the energy of 10 GeV. Circles show the dependence $\theta_{68}(E)$ for the MBC assuming $B$=0.3 fG and the EBL model of \cite{dominguez11}, stars --- $<\theta>(E)$ for the same parameters of the MBC. These two arrays were calculated with the code of \cite{fitoussi17}. Finally, solid line is the analytic estimate of $<\theta>(E)$ from \cite{neronov09}. The agreement between this estimate and the numerical result for $<\theta>(E)$ is reasonable, in agreement with \cite{alvesbatista16}.

Fig.~\ref{fig04} lower-left presents a fit to the observable SED of blazar 1ES 1218+304 ($z$= 0.182) \cite{oikonomou14}--\cite{madhavan13} (circles with statistical uncertainties). Here a simplified version of the Fermi LAT PSF from \cite{kachelriess12} was assumed. EM cascade spectra were calculated with the code of \cite{kachelriess12} assuming the EBL model of \cite{kneiske10}; the approximation method is described in \cite{dzhatdoev17a}.Triangles denote primary (intrinsic) spectrum of the source, stars --- primary (absorbed) observable component, diamonds --- cascade component for the case of $B$= 0, dashed curve --- cascade component for $B(z=0)=10^{-15}$ G, solid curve --- total model fit function. This fit is in qualitative agreement with the observations. Below 100 GeV the suppression of the observable SED (the ``magnetic cutoff effect'') is clearly visible.

\section{Conclusions \label{sec:conclusions}}

In this paper we have considered the basic features of intergalactic EM cascade and presented a new approximation for the $\gamma$-ray mean free path and electron deflection angle in the EGMF. The MBC angular patterns were calculated; it was found that these patterns are practically independent of the range of observable $\gamma$-ray polar angle in the range 0.03-0.3 rad. Finally, we have obtained a fit to the observable SED of 1ES 1218+304 assuming the EGMF strength $B=10^{-15}$ G on 1 Mpc coherence length. In this case the EGMF ``magnetic cutoff effect'' changes the shape of the model spectrum at $E<$100 GeV.

\section*{Acknowledgements}
We are grateful to Prof. I.S. Veselovsky, to the members of the Tokyo University $\gamma$-ray group, especially to Prof. M. Teshima, and to the participants of the Moriond-2017, ISCRA-2017, and ICRC-2017 conferences for many stimulating discussions. We are indebted to Dr. S. Takahashi and Dr. N. Topchiev for providing us with the tables of the GRAINE and GAMMA-400 instruments' angular resolution, Dr. V. V. Kalegaev for permission to use the SINP MSU space monitoring data center computer cluster and to M.D. Nguen for technical support. This work was supported by the Students and Researchers Exchange Program in Sciences (STEPS), the Re-Inventing Japan Project, JSPS.

\newpage
\newpage
\newpage
\newpage
\newpage
\end{document}